\documentclass[12pt,times,twocolumn]{article}
\pdfoutput=1

\usepackage{sbcm}

\usepackage{graphicx,url}
\usepackage[utf8]{inputenc}
\usepackage[hidelinks]{hyperref}
\usepackage{amsmath,amssymb}
\usepackage{verbatim}
\usepackage{siunitx}

\title{Sounderfeit: Cloning a Physical Model with Conditional
  Adversarial Autoencoders}

\author{Stephen Sinclair\inst{1}}

\address{Inria Chile \\
         Av. Apoquindo 2827, piso 12
         Las Condas -- Santiago, Chile
         \email{stephen.sinclair@inria.cl}
}

\newcommand{\todo}[1]{\textbf{todo} \emph{\color{blue} #1}}

\begin{document}
\maketitle

\begin{abstract}
An adversarial autoencoder conditioned on known parameters of a
physical modeling bowed string synthesizer is evaluated for use in
parameter estimation and resynthesis tasks. Latent dimensions are
provided to capture variance not explained by the conditional
parameters. Results are compared with and without the adversarial
training, and a system capable of ``copying'' a given parameter-signal
bidirectional relationship is examined.  A real-time synthesis system
built on a generative, conditioned and regularized neural network is
presented, allowing to construct engaging sound synthesizers based
purely on recorded data.
\end{abstract}

\section{Introduction}




This paper explores the use of an \emph{autoencoder} to mimic the
bidirectional parameter-data relationship of an audio synthesizer,
effectively ``cloning'' its operation.

The \emph{autoencoder} is an artificial neural network (ANN)
configuration in which the network weights are trained to minimize the
difference between input and output, in essense learning the identity
function.
When forced through a bottleneck layer of few parameters, the network
is made to represent the data with a low-dimensional ``code,''
which we call the latent parameters.

Recently adversarial configurations have been proposed as a method of
regularizing this latent parameter space in order to match it to
a given distribution \cite{makhzani2015adversarial}.
The advantages are two-fold: to ensure the available range is
uniformly covered, making it a useful interpolation space; and to
maximally reduce correlation between parameters, encouraging them to
represent orthogonal aspects of the variance.
For example, in a face-generator model, this could translate to
parameters for hair style and the presence of glasses
\cite{radford2015unsupervised}.

Meanwhile, it has also been shown that a generative network can be
conditioned on known parameters \cite{gauthier2014conditional}, to
make it possible to control the output, for example, to generate a
known digit class when trained on MNIST digits.

In this work, these two concepts are combined to explore whether an
adversarial autoencoder can be conditioned on known parameters for use
in both parameter estimation and resynthesis tasks.
In essence, we seek to have the network simultaneously learn to mimic
the transfer function from parameters to data of a periodic signal, as
well as from data to parameters, using adversarial training to
regularize the distribution of the latent space.
Latent dimensions are provided to the network to capture variance not
explained by the conditional parameters, usually refered to in the
image synthesis literature as ``style''; in audio, they may represent
internal state, stochastic sources of variance, or unrepresented
parameters e.g.\ low-frequency oscillators.

Results are visualized and some preliminary evaluations are performed.
Most problems came from issues with the dataset rather than with the
network architecture or training algorithms, and we conclude with some
lessons learned in the art of ``synth cloning'' and how to handle
sampling.
A real-time synthesis system, Sounderfeit, built on a generative,
regularized neural network is presented, allowing to construct
engaging sound synthesizers based purely on recorded data, with
optional conditioning on prior knowledge of parameters.

\section{Configuration}

\section{Dataset}

Given a network with sufficient capacity we can encode any functional
relationship, but for the experiments described herein a periodic
signal specified by a small number of parameters was sought that
nonetheless features some complexity and is related to sound
synthesis.
Thus, a physical modeling synthesizer proved a good choice.
We used the bowed string model from the \emph{STK Synthesis Toolkit in
  C++} \cite{CookS99}, which uses digital waveguide synthesis and is
controlled by 4 parameters: \emph{bow pressure}, the force of the bow
on the string; \emph{bow velocity}, the velocity of the bow across the
string; \emph{bow position}, the distance of the string-bow
intersection from the bridge; and \emph{frequency}, which controls the
length of the delay lines and thus the tuning of the instrument.

The parameters are represented in STK as values from 0 to 128, and
thus we do not worry about physical units in this paper; all
parameters were linearly scaled to a range $[-1,1]$ for input to the
neural network.
The data was similarly scaled for input, and a linear descaling of the
output is performed for the diagrams in this paper.
Additionally, the per-element mean and standard deviations across the
entire dataset were subtracted and divided respectively in order to
ensure similar variance for each discrete step of the waveform period.

To extract the data, a program was written to evaluate the bowed
string model at 48000~kHz for 1~sec for each combination of \emph{bow
  position} and \emph{bow pressure} for integers 0 to 128.
The \emph{bow velocity} and \emph{volume} parameters were both held at
a value of 100.
%
%
For each instance, the last two periods of oscillation were kept, and
since some parameter combinations did not give rise to stable
oscillation, recordings with an RMS output lower than $10^{-5}$ (in
normalized units) over this span were rejected, giving a total of
15731 recordings evenly distributed over the parameter range.
The frequency was selected at 476.5~Hz to count 201 samples to capture
two periods---some parameter combinations changed the tuning slightly,
but inspection by eye of 50 periods concatenated end to end showed
minimal deviation at this frequency for a wide variety of parameters.
Two periods were recorded in order to minimize the impact of any
possible reproduction artifacts at the edges of the recording during
overlap-add synthesis.
The recordings were phase-aligned using a cross-correlation analysis
with a representative random sample, then differentiated by
first-order difference, and 200 sample-to-sample differences were thus
used as the training data, normalized as stated above.
Reproduction thus consists of de-normalizing, concatenating (using
overlap-add with a Blackman window) and first-order integrating the
final signal.
This dataset we refer to as \emph{bowed1}.

Recently-published similar work recommends using log-spectrum
representations rather than raw audio \cite{engel2017neural}, however
we found that since we are concentrating on small two-period,
phase-aligned samples featuring relatively small amount of variance
(as compared to trying to learn a large dataset of fully mixed
instruments), learning the raw audio signal was no problem.
In future work it is possible that different/better results could be
had by using a log-spectrum representation, but in this manner we
avoided the need to perform phase reconstruction.
The use of a differentiated representation also helped to suppress
noise.

As will be discussed below, parameter estimation on new data was not
successful based on this dataset.
To resolve this, a second extended dataset, \emph{bowed2}, was created
in a similar manner, however instead of recording only the steady
state portion, the synthesizer was executed continuously while
changing the parameters randomly at random intervals.
This allowed to capture more dynamic regimes.
100,000 samples uniformly covering the parameter range were captured
for \emph{bowed2}.

\subsection{Learned conditional autoencoding}

While the principle job of the autoencoder is to reproduce the input
as exactly as possible, in this work we also wish to estimate the
parameters used to generate the data.
Thus we additionally \emph{condition} part of the latent space by
adding a loss related to the parameter reconstruction.
This is somewhat different to providing conditional parameters to the
\emph{input} of the encoder
\cite{makhzani2015adversarial,gauthier2014conditional}.

Note that the presence of the latent parameters is what allows for the
fact that we do not assume that the signal is purely deterministic in
the known parameters.
For instance, in a physical signal there maybe internal state
variables that are not taken into account in the initial conditions,
or acoustic characteristics such as room reverb that are not
considered a priori.
Naturally, the less deterministic the signal is in the known
parameters, the more must be left to latent parameters, and the poorer
a job we can expect the parameter reconstruction to do.

\subsection{Generative adversarial regularization}

The code used in the middle layer of an autoencoder, called the
\emph{latent parameters}, which we shall refer to as $z$, when trained
to encode the data distribution $p(x)$, has conditional posterior
probability distribution $q(z|x)$.
As mentioned, it is in general useful to regularize $q(z|x)$ to match
a desired distribution.

Several methods exist for this purpose: a \emph{variational
  autoencoder} (VAE) uses the Kullback-Leibler divergence from a given
prior distribution.
Other measures of difference from a prior are possible.
The use of an adversarial configuration has been proposed
\cite{makhzani2015adversarial} to regularize $q(z)$ based on the
negative log likelihood from a discriminator on $z$.

%
With adversarial regularization, a discriminator is used to judge
whether a posterior distribution $q(z)$ was likely produced by the
generator and is thus sampled from $q(z|x)$, or rather sampled from an
example distribution $p(z)$, which is often set to a normal or uniform
distribution.
%
The discriminator is itself an ANN which outputs a 1 for ``real''
samples of $p(z)$ or a 0 for ``fake'' samples of $p(z)=q(z|x)$.  The
training loss of the generator, which is also the encoder of the
autoencoder, maximizes the probability of fooling the discriminator
into thinking it is a real sample of $p(z)$, while the discriminator
simultaneously tries to increase its accuracy at distinguishing
samples from $p(z)$ and samples from $q(z|x)$.
Thus thus encoder eventually generates posterior $q(z|x)$ to be
similar to $p(z)$.

\subsection{System Architecture Summary}

Putting together the above concepts, the system is composed of two
neural networks and three training steps.

First, the autoencoder network is composed of the encoder $E=f(x)$ and
the decoder/generator $G=g(z,y)$. The discriminator is designed
analogously as $D=h(z)$.
For notational convenience, we also define $G_E=g(E)=g(f(\bold{x}))$,
$D_E=h(E_z)$, and $D_z=h(\bold{z})$ where $\bold{x}=\bold{x}_1\ldots
\bold{x}_s$ are sampled from $p(x)$, $\bold{z}=\bold{z}_1\ldots
\bold{z}_s$ is sampled from $p(z)$, and $s$ is the batch size.
$E_z(x)$ and $E_y(x)$ are the first $n$ and the last $m$ dimensions of
$f(x) \in [z_1 \cdots z_n \;\; y_1 \cdots y_m]$, respectively.
In the current work, $f(x)$ and $g(z,y)$ are simple one-hidden-layer
ANNs with one non-linearity $\zeta$ and linear outputs:
\begin{align}
  f(x)=\zeta ( x \cdot w_1 + b_1 ) \cdot w_2 + b_2 \\
  g(z)=\zeta ( z \cdot w_3 + b_3 ) \cdot w_4 + b_4 \\
  h(z)=\zeta ( z \cdot w_5 + b_5 ) \cdot w_6 + b_6
\end{align}
We used the rectified linear unit $\zeta(x)=\max(0,x)$, but had
similar results with $\tanh$ non-linearities.

The data, described below, was composed of 200-wide 1-D vectors, and
we had acceptable results using hidden layers of half that size, so
$w_1 \in \mathbb{R}^{200 \times 100}$, $w_2 \in \mathbb{R}^{100 \times
  {(n+m)}}$ and $w_3,w_5 \in \mathbb{R}^{{(n+m)}\times 100}$, $w_4,w_6
\in \mathbb{R}^{100\times 1}$, where ${(n+m)}$ was 2 or 3, depending on
the experiment.  The bias vectors $b_1 \ldots b_6$ had corresponding
sizes accordingly.
Hyperparameter random search was used to guide, but not automatically
select hyperparameters.

The training steps were performed in the following order for each batch:
\begin{enumerate}
\item A stochastic gradient descent (SGD) optimiser with a learning
  rate of 0.005 was used to train the full set of autoencoder weights
  $w_1 \ldots w_4$, and $b_1 \ldots b_4$, minimizing both the data $x$
  reconstruction loss and parameter $y$ reconstruction loss,
  $\mathcal{L}_\textrm{AE}$ by back-propagation.
  The weighting parameter $\lambda=0.5$ is described below.
\item SGD with learning rate 0.05 was used to train the generator
  weights and biases $w_1$, $w_2$, $b_1$, and $b_2$.
  The negative log-likelihood $\mathcal{L}_\textrm{G}$ was minimized
  by back-propagation.
\item SGD with learning rate 0.05 was used to train the discriminator
  weights and biases $w_5$, $w_6$, $b_5$, and $b_6$.
  The negative log-likeli\-hood $\mathcal{L}_\textrm{D}$ was
  minimized by back-propagation.
\end{enumerate}

where,
\begin{align}
  \mathcal{L}_\textrm{AE}&=\sum(\bold{x}-g(f(\bold{x})))^2
  + \lambda\sum(\bold{y} - g(\bold{x}))^2 \\
  \mathcal{L}_\textrm{G}&=-\sum\log(D_E) \\
  \mathcal{L}_\textrm{D}&=-\sum(\log(D_z) + \log(1-D_E).
\end{align}

Experiments were performed using the TensorFlow framework
\cite{tensorflow2015-whitepaper}, which implemented the
differentiation and gradient descent (back-propagation) algorithms.
A small batch size of 50 was used, with each experiment evaluated
after 4,000 batches.  It was found that smaller batch sizes worked
better for the adversarial configuration, since the updates of each
step are interleaved.
Matrices $\bold{z}$ and $\bold{x},\bold{y}$ were sampled independently
from $\bold{Z}\sim{}p(z)=\mathcal{U}(-1,1)$ and
$(\bold{X},\bold{Y})\sim{}p(x,y)$ for each step, where
$\mathcal{U}(a,b)$ is the uniform distribution in range $[a,b]$
inclusive.

\section{Experiments}

Six conditions were tested in order to explore the role of
conditional and latent parameters.
The number of known parameters in the dataset was 2.  We tried
training the \emph{bowed1} dataset with and without an extra latent
parameter.  We label these conditions $D1_Z2_Y$ and $D0_Z2_Y$
respectively.
The third condition, $N1_Z2_Y$, was like the $D1_Z2_Y$ condition but
without adversarial regularization on $q(z|x)$.

In order to further understand how latent parameters may help capture
unknown variance, we tested a condition of treating \emph{bow
  pressure} as a known parameter and leaving \emph{bow position} to be
captured by $z$.
This was accomplished simply by repeating $D1_Z2_Y$ with one missing
$y$ parameter, thus labeled $D1_Z1_Y$.

Finally, to compare conditioning with the ``natural'' distribution of
the data among latent parameters and the effects of adversarial
regularization thereupon, two configurations with no conditonal
parameters, with and without the discriminator, were explored, named
$D2_Z0_Y$ and $N2_Z0_Y$ respectively.

\section{Results}

\begin{figure}
  \centerline{\includegraphics[width=0.5\textwidth]{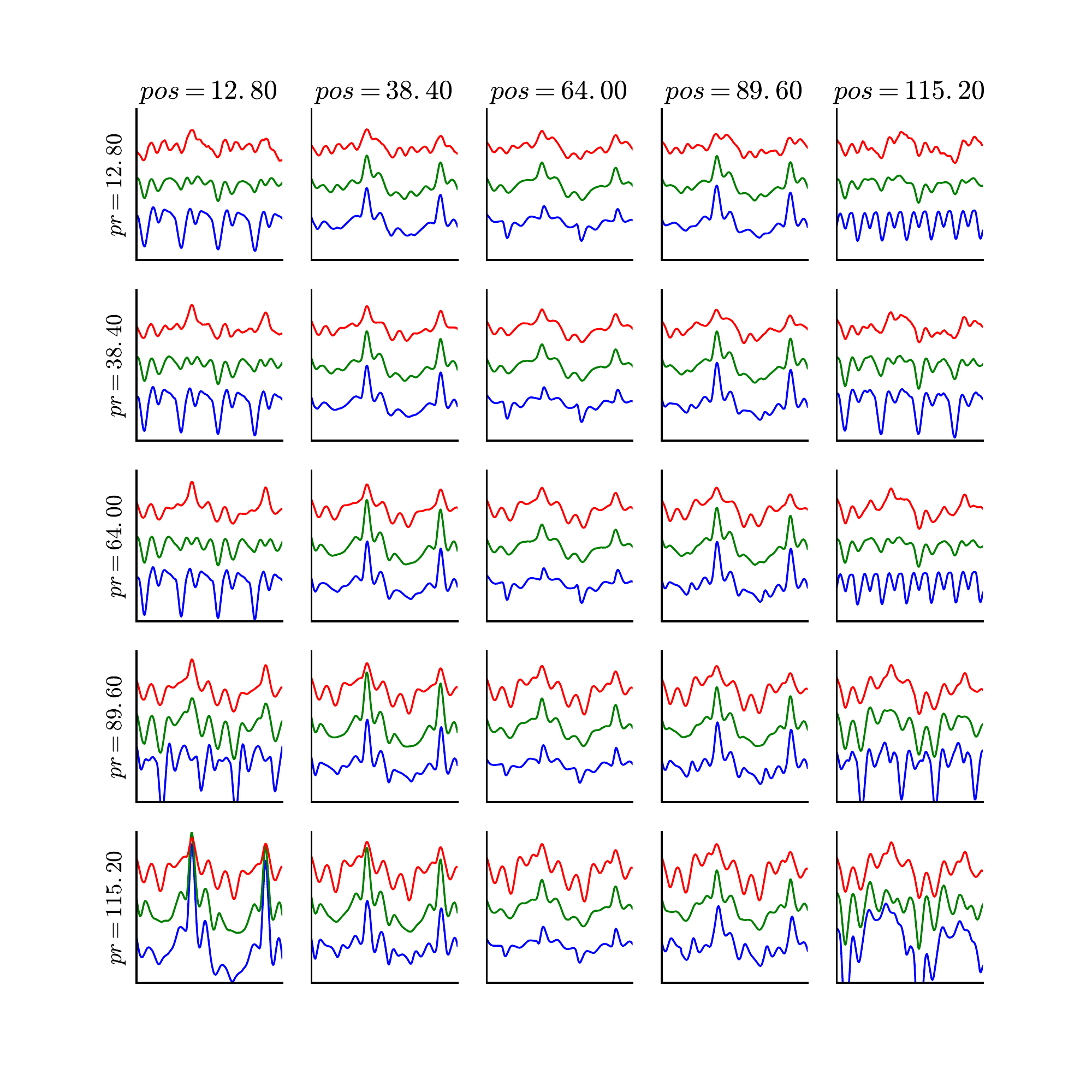}}
  \caption{ Output of $D1_Z2_Y$ as \textit{pressure} $y_0$ and
    \textit{position} $y_1$ are changed, with $z_0=0$.  Top (red) is
    the decoder, middle (green) is the autoencoder, bottom (blue) is
    the dataset.
  }
  \label{bowed-D1Z2Y-varypv}
\end{figure}

Figure~\ref{bowed-D1Z2Y-varypv} demonstrates the results of $D1_Z2_Y$.
Comparing the middle and bottom curves, we can see that while it has
some trouble with low values of \emph{bow pressure} and the extremes
of \emph{bow position}, the autoencoder is able to more or less encode
the distribution in our dataset.
The top curve (red) was generated by explicitly specifying the same
parameters instead of letting the autoencoder infer them, and
demonstrates the output for parameter-driven reconstruction if $z_0$
is held constant.
Although not a perfect reproduction, this demonstrates that parameters
were conditioned according to the dataset, and thus the ANN models the
data-parameter relationship.

\begin{figure}
  \centerline{
    \includegraphics[width=0.5\textwidth]{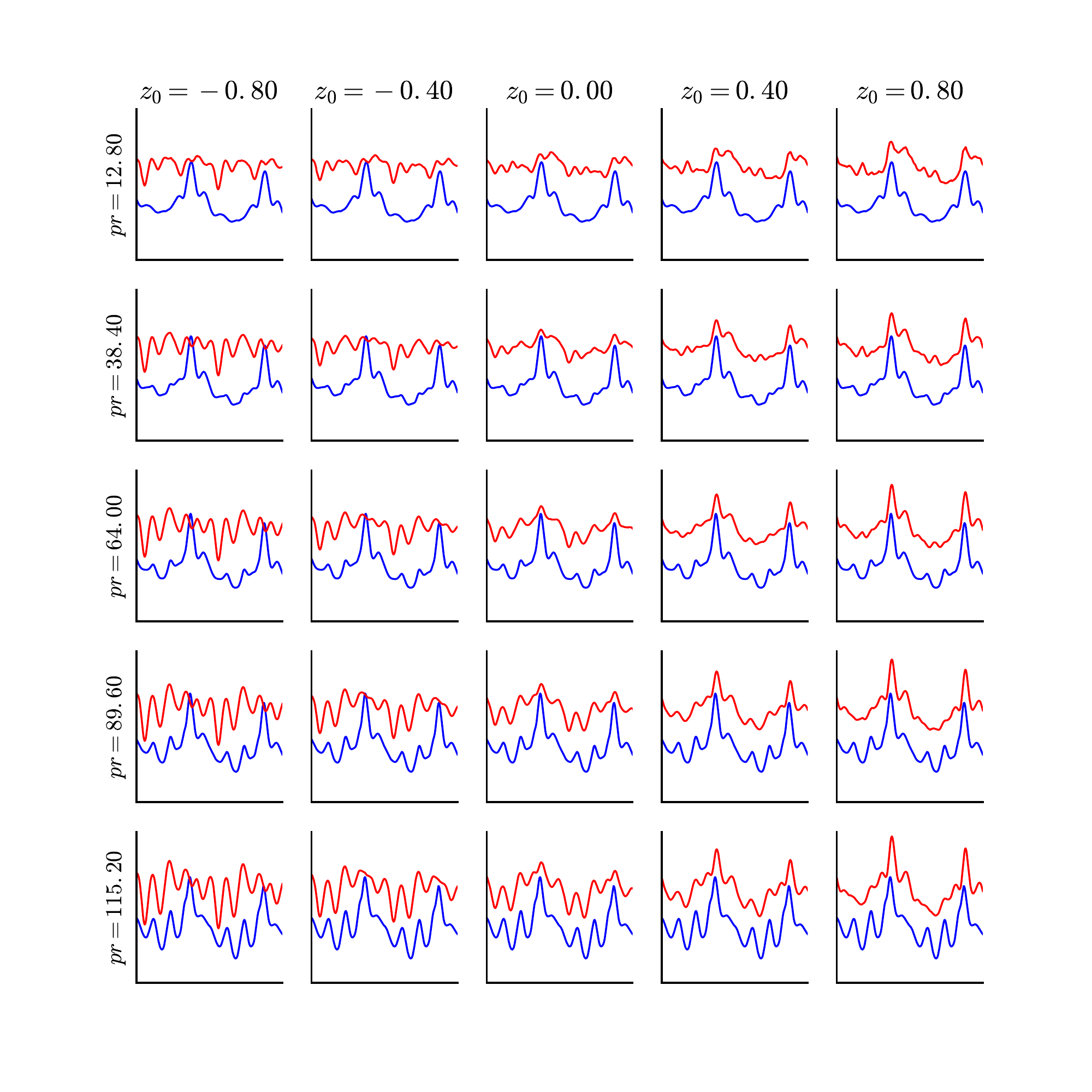}}
  \caption{ Output of $D1_Z2_Y$ as \emph{bow position} is set to 100,
    and \emph{bow pressure} and latent $z_0$ variable are changed. Top
    (red): Decoder output; bottom (blue): dataset. }
  \label{bowed-D1Z2Y-varyvz}
\end{figure}

The role of $z$ is now considered in Figure~\ref{bowed-D1Z2Y-varyvz},
by holding \emph{bow position} constant ($y_1=100$) and examining how
the signal changes with $z_0$.
One notices that for some values of $z_0$ the signal matches well, and
for others it varies from the target signal.
For example, we can see that in this case, high values of $z_0$ push
the signal towards two sharp peaks, while low values of $z_0$ tend
towards more oscillations; both $z_0=-0.8$ and $z_0=0.8$ resemble the
$pr=115.2$ condition, but in different aspects.
Meanwhile there is consistency with the ``stylistic'' influence of
$z_0$ on the signal for different values of bow pressure.

\begin{figure}
  \centerline{
    (a)\includegraphics[width=0.4\textwidth]{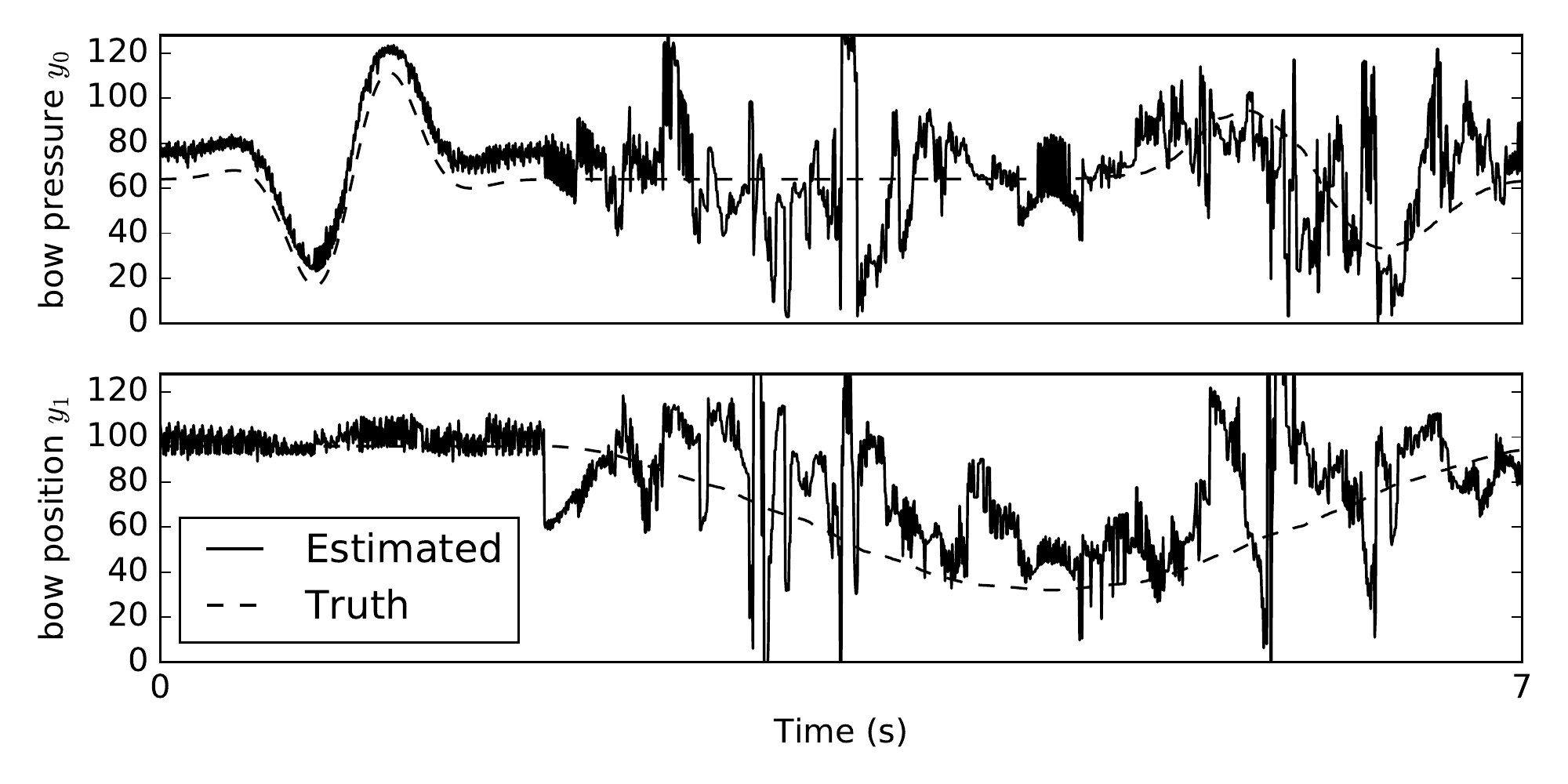}
    }
  \centerline{
    (b)\includegraphics[width=0.4\textwidth]{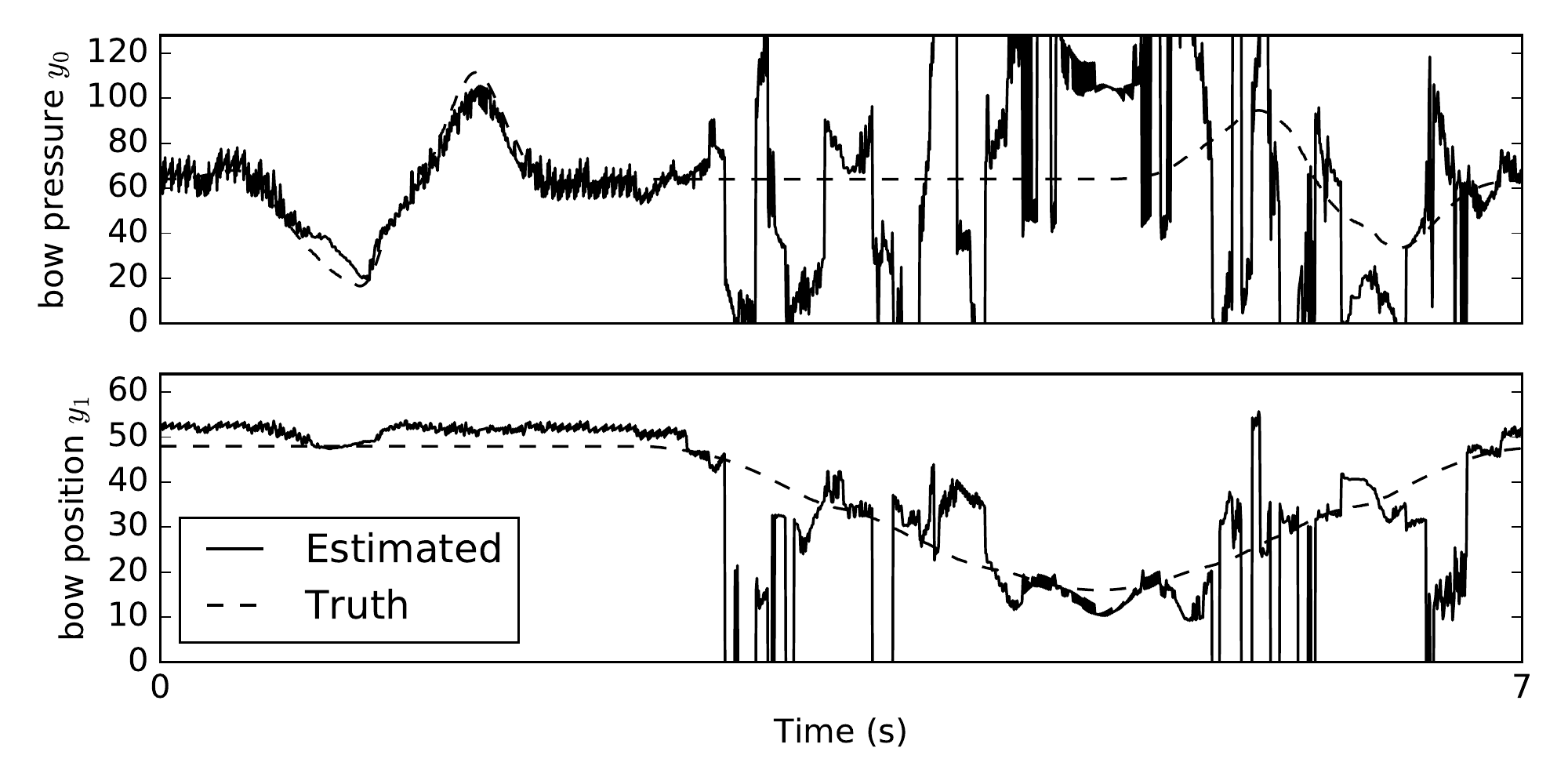}}
  \centerline{
    (c)\includegraphics[width=0.4\textwidth]{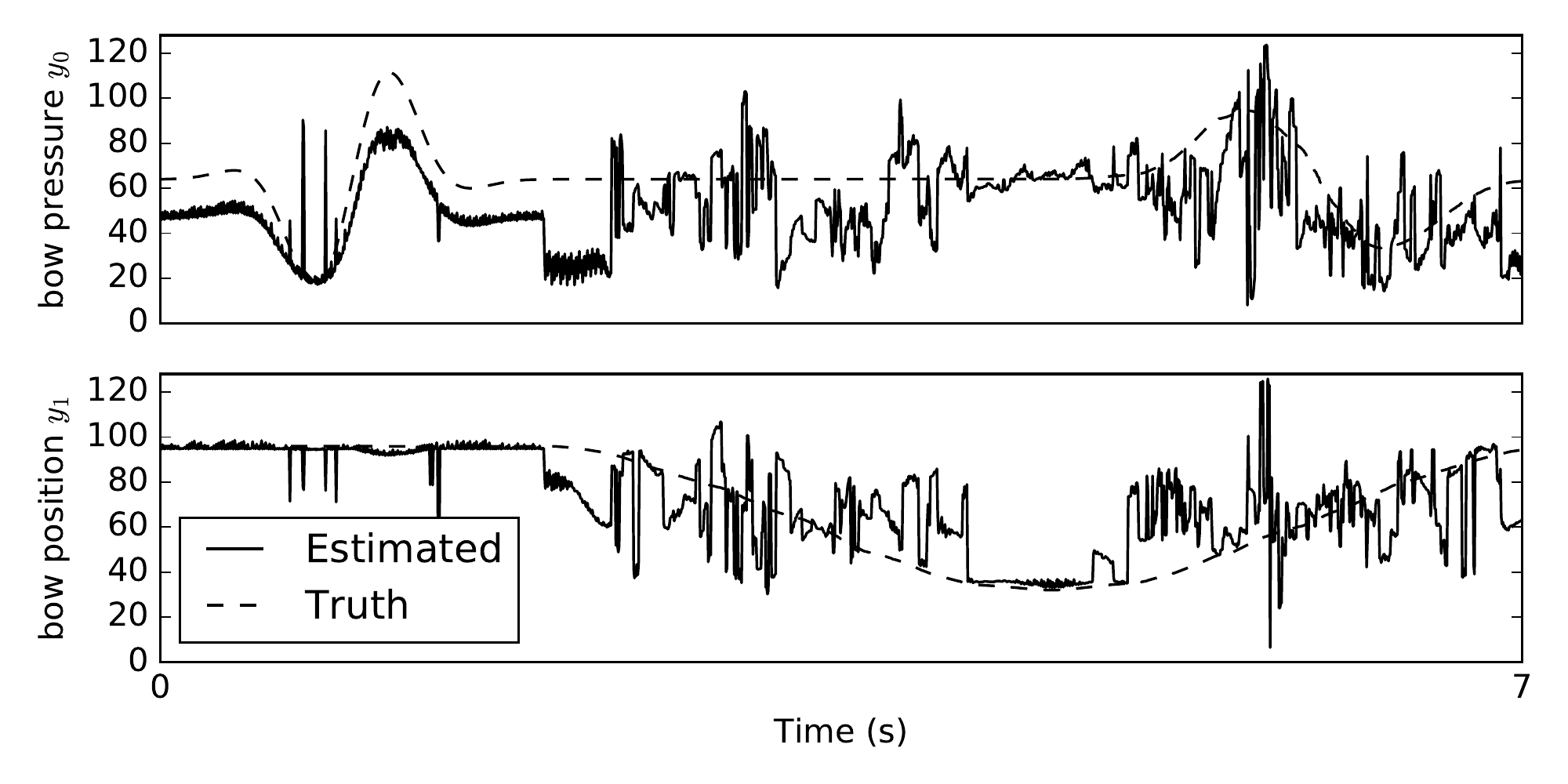}}
  \centerline{
    (d)\includegraphics[width=0.4\textwidth]{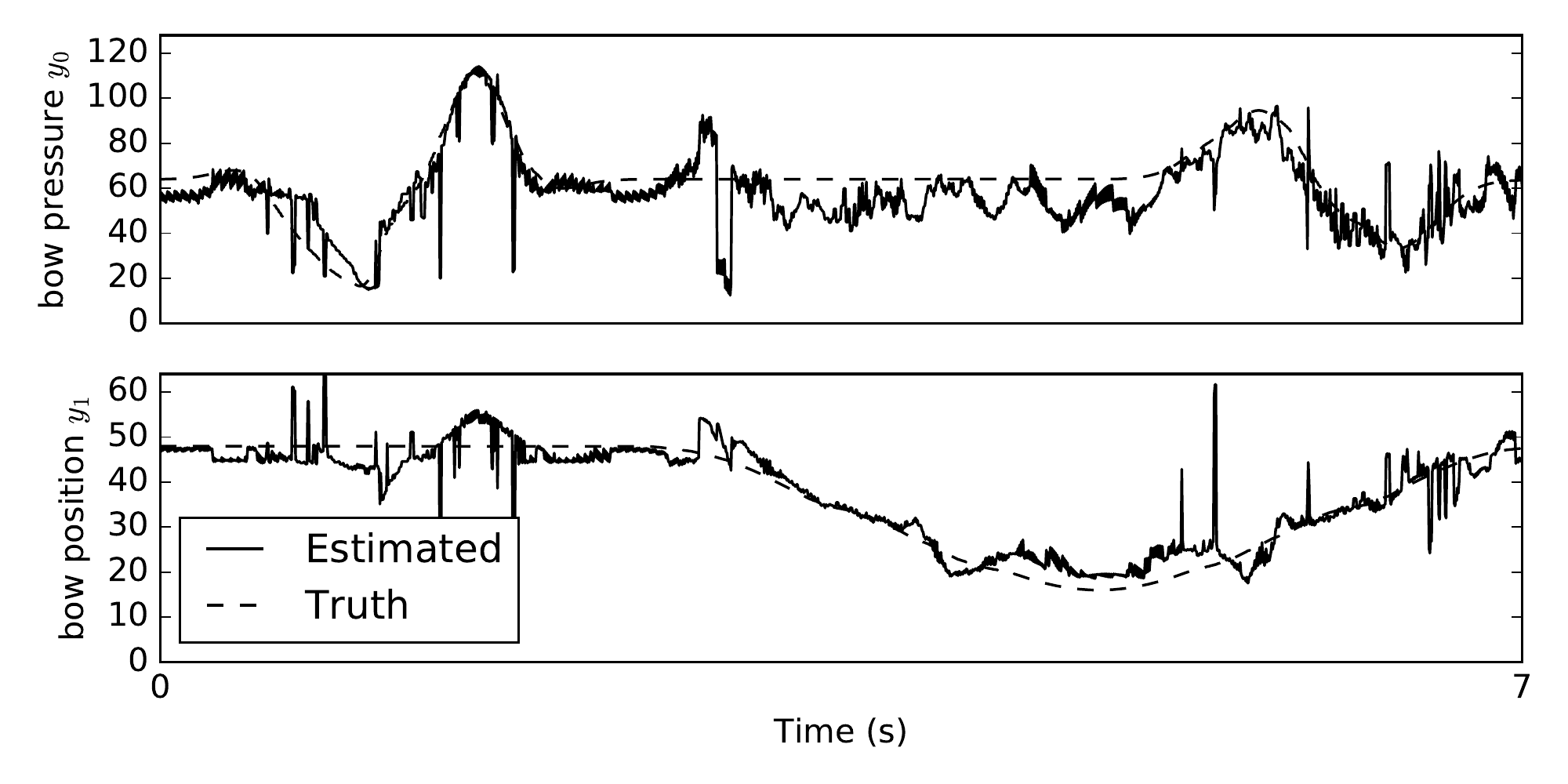}}
  \caption{ Parameter estimation performance of the $D1_Z2_Y$ network
    for (a) \emph{bowed1} full dataset, RMS error=23.23; (b)
    \emph{bowed1} half dataset, RMS error=33.45; (c) \emph{bowed2}
    full dataset, RMS error=19.54; (d) \emph{bowed2} half dataset, RMS
    error=8.61. }
  \label{bowed-known}
\end{figure}

Finally, we look at the encoder (parameter estimator) performance, by
producing a \emph{new} signal from the synthesizer with a parameter
trajectory starting with variation only in \emph{bow pressure} and
then variation only in \emph{bow position}, and then variation in both
parameters.
Figure~\ref{bowed-known}(a) shows actually rather disappointing
performance in this respect, however it does clarify some information
not present in the previous analysis: the estimation is clearly better
for \emph{bow pressure}, but easily disturbed by changes in \emph{bow
  position}.
Nonetheless we see the \emph{tendency} of the estimate in the right
direction, with rather a lot of flipping above and below the center.
Since varying the hyperparameters of our network did not solve this
problem, we hypothesized that this error could come from two sources:
(1) ambiguities in the dataset---indeed, if one examines the shape of
the signal as \emph{bow position} changes, one notices a symmetry
between values on either side of \emph{pos}=64.
By consequence the inverse problem is underspecified, leading to
ambiguity in the parameter estimate.  (2) underrepresented variance in
the dataset; the new testing data varies continuously in the
parameters, but the dataset was constructed based on the per-parameter
steady state.

To investigate this, the network was trained on a ``half dataset'',
consisting only of samples of \emph{bowed1} where \emph{bow position}
$< 64$.
Furthermore, as mentioned, an extended dataset, \emph{bowed2}, was
constructed based on random parameter variations.

Results in Figure~\ref{bowed-known}(b)-(d) show that training on the
half-\emph{bowed1} dataset changed the character of errors, but did
not improve overall, however the extended \emph{bowed2} dataset gave
improved parameter estimation, and much improved in the \emph{bow
  position} $< 64$ case.
Thus it can be concluded that both sources contributed to parameter
estimation difficulties.

\begin{figure}
  \centerline{
    \includegraphics[width=0.4\textwidth]{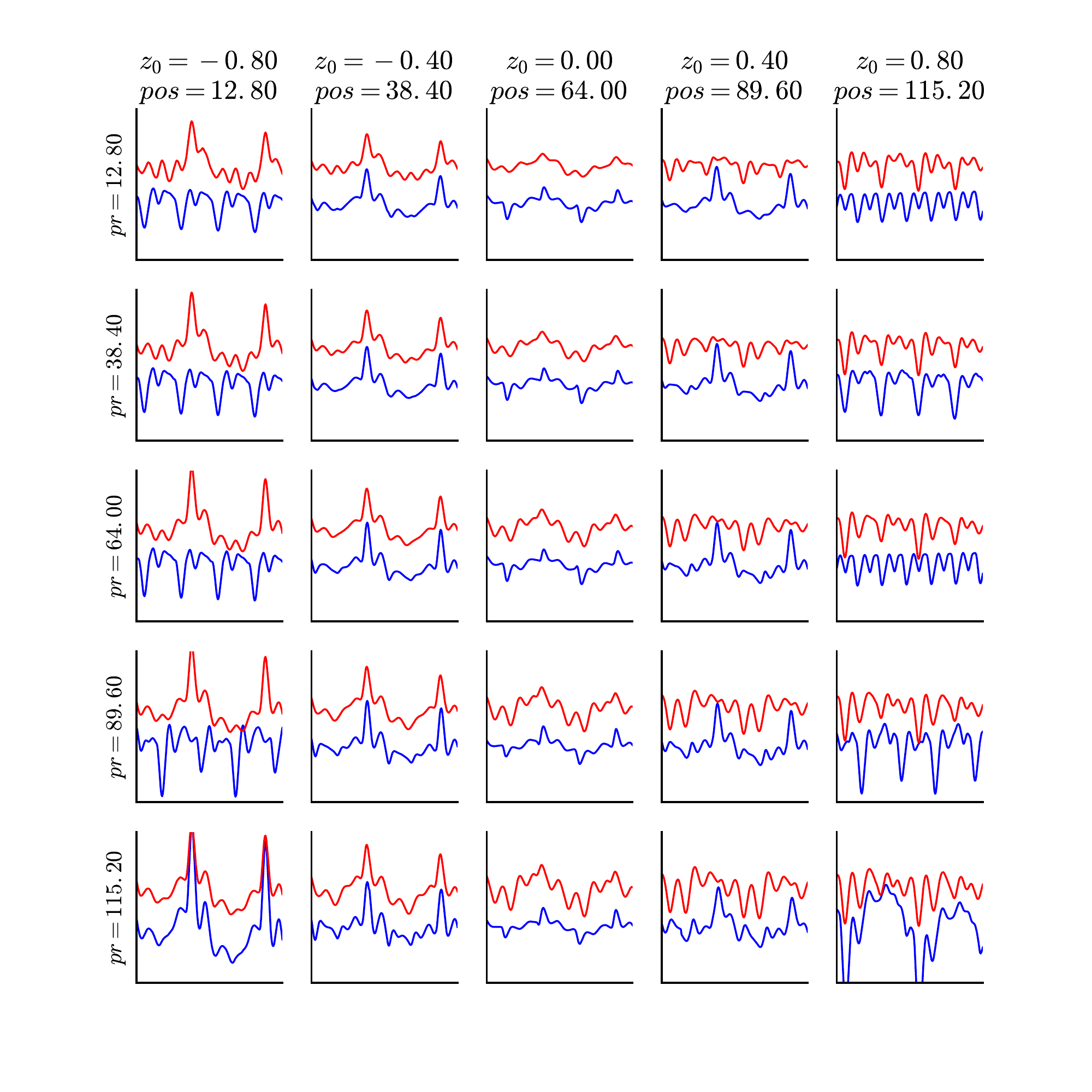}}
  \caption{ Top (red): Output of $D1_Z1_Y$ as \emph{bow pressure}
    $y_0$ and the latent variable $z_0$ are changed; bottom (blue):
    \emph{pr} and \emph{pos} from the dataset. }
  \label{bowed-D1Z1Y-varypv}
\end{figure}

We also examine using $D1_Z1_Y$ how the network performs if only 1
parameter is conditioned.
In Figure~\ref{bowed-D1Z1Y-varypv} it can be seen that the signals
match in many cases, very similar indeed to
Figure~\ref{bowed-D1Z2Y-varyvz}.
However, this is not a given, since the parameter on the horizontal
axis, \emph{bow position}, was not conditioned!
Indeed, it is reflected by the $z_0$ variable automatically, since it
is the principle source of variance unexplained by the conditioned
parameter \emph{bow pressure}.

%

\begin{figure}[t]
  \centerline{
    \includegraphics[width=0.4\textwidth]{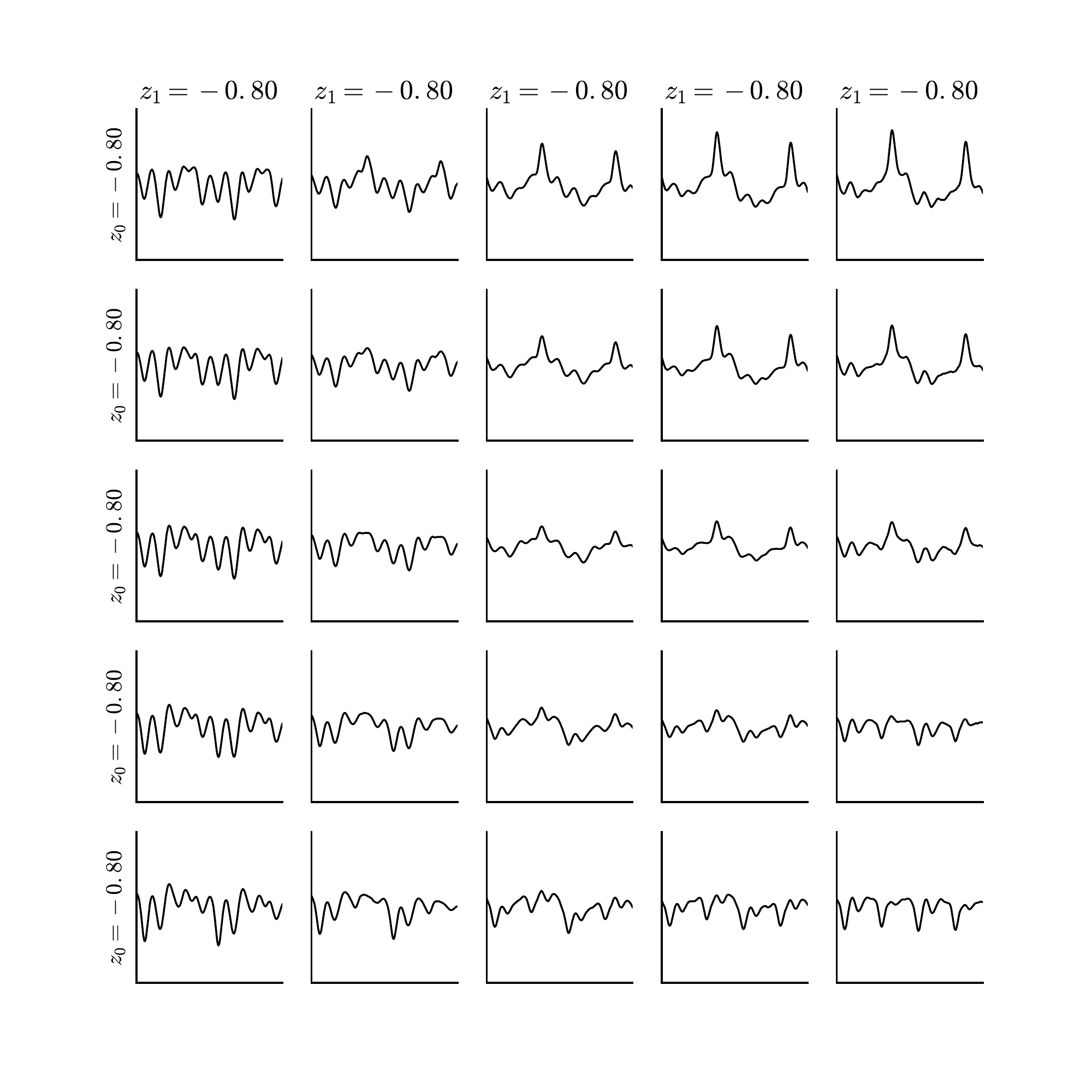}}
  \caption{ Output of $D2_Z0_Y$, varying $z_0$ and $z_1.$
    } 
  \label{bowed-2z0y}
\end{figure}

\begin{figure}[t]
  \centerline{
    \includegraphics[width=0.3\textwidth]{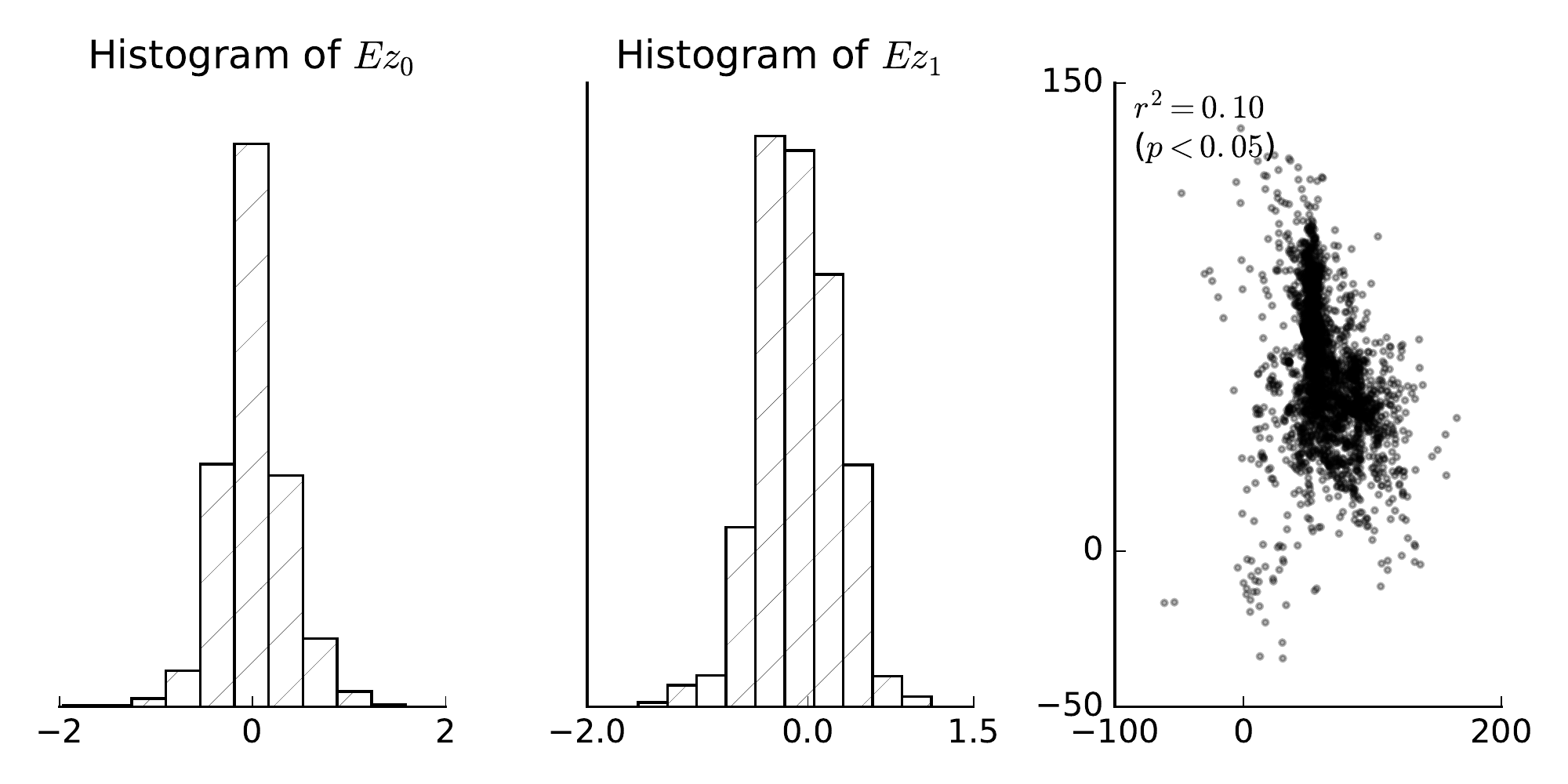}}
  \centerline{
    \includegraphics[width=0.3\textwidth]{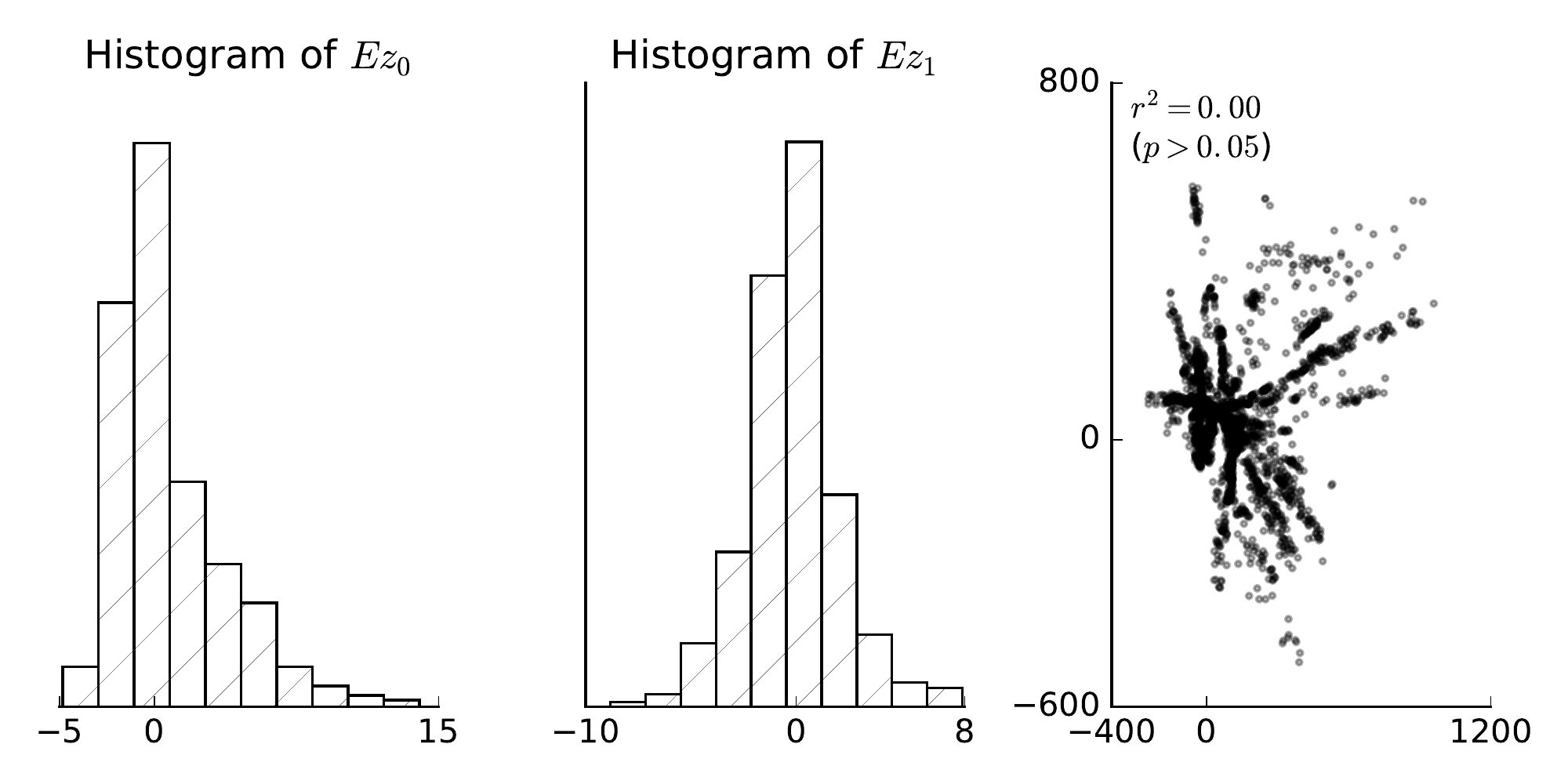}}
  \caption{ Parameter distributions: top, $D2_Z0_Y$ (regularized);
    bottom, $N2_Z0_Y$ (no regularization). }
  \label{bowed-paramdist}
\end{figure}

Figure~\ref{bowed-2z0y} shows the resulting parameter space if both
parameters are left to be absorbed by the unsupervised latent space.
Indeed it seems that with regularization, the system is encouraged
to cover the entire range of variance in the dataset.
Without regularization ($N2_Z0_Y$) we see some relationship between
the two inferred variables $z_0$ and $z_1$
(Fig.~\ref{bowed-paramdist})---although it appears more complex than
could be captured by a Pearson's correlation---while this is
completely gone for the regularized version ($D2_Z0_y$).
%
The star shape is generated because without regularization, the
autoencoder attempts to maximally separate various aspects of the
variance in a reduced 2-dimensional space, which can be useful for
data analysis but does not produce a good interpolation space.

\begin{figure}[t]
  \centerline{
    \includegraphics[width=0.5\textwidth]{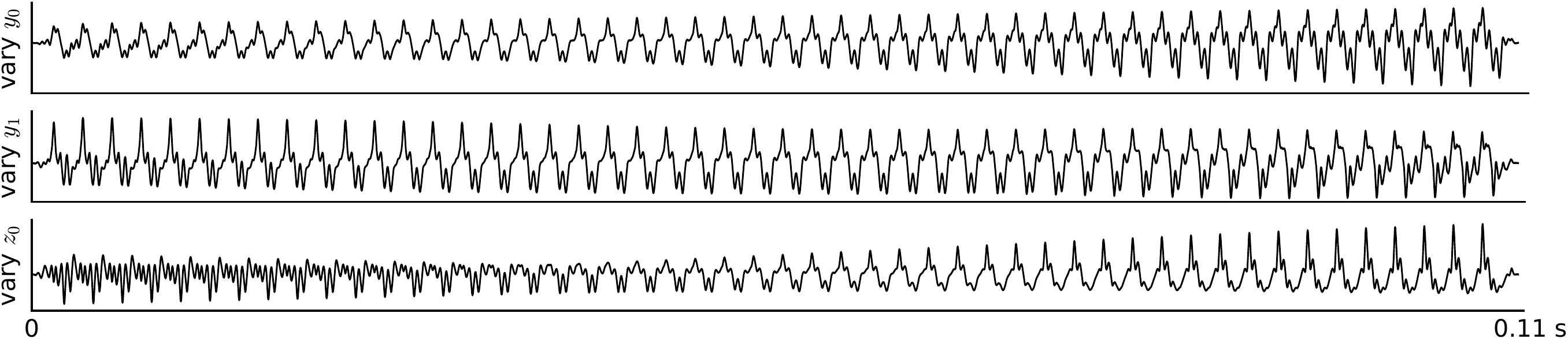}}
  \caption{ Overlap-add output of $D1_Z2_Y$, varying each parameter
    over a short interval. }
  \label{bowed-sound}
\end{figure}

Finally, we found that with this small decoder network of $100\times3$
weights and 100 biases, an overlap-add synthesis could be performed in
real time on a laptop computer (10 seconds took 8.5 seconds to
generate), and
we can thus create a data-driven wavetable synthesizer, which we call
Sounderfeit, with a number of adjustable parameters.
The regularization encourages the parameter space to be
``interesting,'' in the sense that they represent orthogonal axes
within the distribution that cover a defined range and tend towards
uniform coverage without ``holes.''  This is demonstrated in
Figure~\ref{bowed-sound}.


\section{Conclusions}

These experiments showed some modest success in copying the
parameter-data relationship of a physical modeling synthesizer.

Like many machine learning approaches, the quality of results depend
strongly on the hyperparameters used: network size and architecture,
learning rates, conditional regularization weights, etc., and these
must be adapted to the dataset.
Shown are results from the best parameters found after some
combination of automatic and manual optimisation on this specific
dataset, which we use to demonstrate some principles
of the design, however it should be noted that actual results varied
sometimes unexpectedly with small changes to these parameters.
This hyperparameter optimization is non-trivial, especially when it
comes to audio where mean squared error may not reveal much about the
perceptual quality of the results, and so a lot of trial and error is
the game.
Thus, a truly ``universal'', turn-key synthesizer copier would require
future work on measuring a combined hypercost that balances well the
desire for good reproduction with good parameter estimation quality,
and well-distributed latent parameters.  Such work could go beyond
mean squared error to involve perceptual models of sound perception.

Some practical notes: (1) We found that getting the adversarial method
to properly regularize the latent variables in the presence of
conditional variables is somewhat tricky; the batch size and relative
learning rates played a lot in balancing the generator and
discriminator performances.  New research in adversarial methods is a
current area of investigation in the ML community and many new
techniques could apply here; moreover comparison with variational
methods is needed.
(2) Expectedly, we found the parameter estimation extremely sensitive
to phase alignment; we tried randomizing phase of examples during
training, which gave better parameter estimates, but this was quite
damaging to the autoencoder performance.  In general oversensitivity
to phase is a problem with this method, a downside to the time domain
representation.

Nevertheless we have attempted to outline some potential for use of
autoencoders and their latent spaces for audio analysis and synthesis
based on a specific signal source.
Only a very simple fully-connected single-layer architecture was
explored; myriad improvements could likely be made using convolutional
layers, different activation functions, etc.
More important than the quality of these specific results, we wish to
point out the modular approach that autoencoders enable in modeling
oscillator periods of known and unknown parameters, and that, in
contrast to larger datasets covering many instruments
\cite{engel2017neural}, interesting insights can be had even on small
data.



The motivation of the work could be questioned, in the sense that a
black box model seemingly does not bring much to the table in the
presence of an existing, semantically-rich physical model.
Indeed, in this work a digital synthesizer was used as an easy way to
gain access to a fairly complicated but clean signal with a small
number of parameters.
In principle this method could be used on much richer, real instrument
recordings, provided that a measurement or estimate of
acoustically-relevant parameters is available.
For example, we have applied it to recorded vowel vocalizations, with
the vowel category as a single continuous variable, creating a
real-time vowel synthesizer similar to the bowed string results with a
controllable vowel knob and other characteristics represented by the
latent space.

Another question may be why perform simultaneous estimation and
generation with the same network.
Indeed, part of the long-term goals are to ``play'' somewhat more with
the latent space, such as using it for what is known in the audio
community as cross-synthesis, or in the machine learning community as
``style transfer'', i.e., swapping the bottom and top halfs of two
such autoencoder networks, allowing to drive a synthesizer by both
conditioned and latent parameters estimated on an incoming signal.

\bibliographystyle{unsrt}
\bibliography{autoenc-bowed}

\end{document}